\documentclass[20pt]{article}
\usepackage{amsmath}
\usepackage{amssymb}
\usepackage{latexsym}
\usepackage{amsmath}
\usepackage{lineno}
\usepackage{graphicx}
\usepackage{float}
\usepackage{subfigure}
\usepackage{geometry}
\usepackage{graphics}
\usepackage{subfigure}
\usepackage{graphicx}
\usepackage{multirow}
\usepackage{booktabs}
\usepackage{hyperref}
\usepackage{cite}
\makeatletter

\newcommand{\be}{\begin{equation}}
\newcommand{\ee}{\end{equation}}

\begin{document}
\begin{center}
\large {\bf Hawking radiation of massive bosons via tunneling from black strings}
\end{center}
\begin{center}
Zhong-Wen Feng
 $\footnote{E-mail:zwfengphy@163.com}$
Shu-Zheng Yang
\end{center}

\begin{center}
\textit{College of Physics and Space Science, China West Normal University, Nanchong 637009, China}
\end{center}

\noindent
{\bf Abstract:} In the present paper, the Hawking radiation of massive bosons from 4-dimensional and 5-dimensional black strings are studied in quantum tunneling formalism. Firstly, we derive the Hamilton-Jacobi equation set via the Proca equation and WKB approximation. Then, the tunneling rates and Hawking temperature of the black strings are obtained. Our calculations show that the tunneling rates and Hawking temperatures are related to the properties of black strings' spacetime. When compare our results with those of scalars and fermions cases, it finds that they are the same.

\noindent
{\bf Keywords:}Massive bosons;  Hawking radiation; Black strings

\section{Introduction}	
\label{Int}
One common feature among black hole physics is that the existence of black holes' thermal properties.  This idea was first proposed by Bekenstein, who  proved that the entropy of black holes have satisfies the relation $S = {A \mathord{\left/ {\vphantom {A 4}} \right. \kern-\nulldelimiterspace} 4}$, where $A$ is the horizon area \cite{ch1}. In 1976, inspired by this entropy theory of black holes, Hawking discovered that black holes have thermal radiation (now people called this theory as Hawking radiation), and the temperature of black holes (Hawking temperature) can be expressed as $T = {\kappa  \mathord{\left/ {\vphantom {\kappa  {2\pi }}} \right. \kern-\nulldelimiterspace} {2\pi }}$ with the surface gravity of black holes $\kappa$ \cite{ch2}. The Hawking radiation is very for the foundations of physics, which was great influence on gravitational theory, quantum mechanism, and thermodynamic. Therefore, the Hawking radiation has received wide attention.

Based on the original theory of Hawking, the black hole radiation was considered as a quantum tunneling effect near the horizon. So, people can discuss the Hawking radiation and the temperature of black holes via the quantum tunneling method. In Ref.~\cite{ch3}, Parikh and Wilczek put forward the first kind of quantum tunneling method, which is called as the Null Geodesic method. Using this method, they calculate the tunneling behaviors of massless scalar particles from Schwarzschild black hole as well as Reissner-Nordstr\"{o}m black hole and obtained those black holes' temperature. The Hamilton-Jacobi ansatz is another kind of quantum tunneling method. In Ref.~\cite{ch3+}, the authors studied the massive scalar particles tunneling from spherically symmetric spacetimes.  Subsequently, Kerner and Mann extended Hamilton-Jacobi ansatz to fermions case and studied fermion tunneling. They first applied the WKB approximation to Dirac equation. Then, neglecting the higher-order terms of $\mathcal{O(\hbar)}$, the resulting equation to leading order in $\hbar$ becomes the Hamilton-Jacobi equation. According the Hamilton-Jacobi equation, Kerner and Mann investigated the Hawking radiation of fermions via quantum tunneling from in the Rindler spacetime and a general non-rotating black hole \cite{ch4,ch5}. Soon after that, Banerjee and Majhi developed the Hamilton-Jacobi ansatz to solve the  information loss paradox \cite{ch5+,ch6+}. Nowadays, the radiation behavior in many complicated spacetimes has been investigated via quantum tunneling method \cite{ch6,ch7,ch8,ch9,ch10,ch11,ch12,ch13}.

It is a well-known issue that the bosons or vector particles (e.g. $Z$, $W^ \pm$) play a very fundamental role as in the standard model for electroweak interaction and the high energy physics. Therefore, the quantum tunneling of bosons from the black hole has recently attracted people's attention \cite{ch13+,ch14+,ch14,ch15,ch18+,ch19+,ch16,ch17,ch20+,ch21+,ch22+,ch23+}. In Ref.~\cite{ch14,ch15}, Kruglov first used the Proca equation to calculate the Hawking radiation of vector from low dimensional spacetimes. Then, we extended Kruglov's work and studied the tunneling behavior of massive bosons from Kerr-de Sitter black hole and 5-dimensional Schwarzschild-Tangherlini black hole \cite{ch18+}.  On the other hand, the black strings are important solutions of the Einstein-Maxwell equations. Those cylindrically symmetric solutions are helpful for us to study the AdS/CFT correspondence. Thus, in the paper we will use Proca equation to study the vector particles tunneling from 4-dimensional  rotating black string and 5-dimensional  black string.

The remainders of this paper are outlined as follows. Using the Proca equation and WKB approximation, the tunneling process of massive bosons from 4-dimensional rotating black string are investigated in Section~\ref{II}. In Section~\ref{III}, we extend our calculations and study massive bosons Hawing radiation from 5-dimensional static black string. In Section~\ref{Dis}, some conclusions and discussions are presented.

\section{Massive bosons tunneling from the rotating black string}
\label{II}
In Ref.~\cite{ch16}, the authors analyzed the Einstein-Maxwell equations with a cosmological constant, which in a stationary spacetime admitting an isometry group of $R\times \rm{U(1)}$. In their work, the rotating black string metric is given by
\begin{align}
\label{eq1}
 ds^2 & =  - \left[ {\alpha ^2 r^2  - \frac{{4M}}{{\alpha r}}\left( {1 - \frac{{a^2 \alpha ^2 }}{2}} \right)} \right]dt^2  + \left[ {\alpha ^2 r^2  - \frac{{4M}}{{\alpha r}}\left( {1 - \frac{3}{2}a^2 \alpha ^2 } \right)} \right]^{ - 1} dr^2
\nonumber \\
&  + \left( {r^2  - \frac{{4Mra^2 }}{{\alpha r}}} \right)d\phi ^2  + \alpha ^2 r^2 dz^2  - \frac{{8Ma\sqrt {1 - \frac{{a^2 \alpha ^2 }}{2}} }}{{\alpha r}}dtd\phi ,
\end{align}
where $\alpha^2$ is the negative cosmological constant, which satisfy the relation $\alpha  =  - {\Lambda  \mathord{\left/ {\vphantom {\Lambda  3}} \right.
 \kern-\nulldelimiterspace} 3}$. Meanwhile, we denote that $a^2 \alpha ^2  = 1 - {\varepsilon  \mathord{\left/ {\vphantom {\varepsilon  M}} \right. \kern-\nulldelimiterspace} M}$, $\varepsilon  = \sqrt {M^2  - {{8J^2 \alpha ^2 } \mathord{\left/ {\vphantom {{8J^2 \alpha ^2 } 9}} \right. \kern-\nulldelimiterspace} 9}}$, $J = \left( {{{3Ma} \mathord{\left/  {\vphantom {{3Ma} 2}} \right.
 \kern-\nulldelimiterspace} 2}} \right)\sqrt {1 - {{a^2 \alpha ^2 } \mathord{\left/  {\vphantom {{a^2 \alpha ^2 } 2}} \right.  \kern-\nulldelimiterspace} 2}} $, where $M$ and $J$  are the mass and angular momentum densities of the rotating black string, respectively. For simplicity,  the metric (\ref{eq1}) becomes to
\begin{eqnarray}
\label{eq2}
ds^2  =  - \Delta \left( {\gamma dt - \frac{\delta }{{\alpha ^2 }}d\phi } \right)^2  + r^2 \left( {\gamma d\phi  - \delta dt} \right)^2  + \frac{1}{\Delta }dr^2  + \alpha ^2 r^2 dz^2 ,
\end{eqnarray}
with $\Delta  = a^2 r^2  - {b \mathord{\left/ {\vphantom {b {\alpha r}}} \right. \kern-\nulldelimiterspace} {\alpha r}}$, $b = 4M\left( {1 - {{3a^2 \alpha ^2 } \mathord{\left/ {\vphantom {{3a^2 \alpha ^2 } 2}} \right. \kern-\nulldelimiterspace} 2}} \right)$,$\gamma  = \sqrt {{{\left( {2 - a^2 \alpha ^2 } \right)} \mathord{\left/ {\vphantom {{\left( {2 - a^2 \alpha ^2 } \right)} {\left( {2 - 3a^2 \alpha ^2 } \right)}}} \right. \kern-\nulldelimiterspace} {\left( {2 - 3a^2 \alpha ^2 } \right)}}}$ ,${{a\alpha ^2 } \mathord{\left/ {\vphantom {{a\alpha ^2 } {\sqrt {1 - {{3a^2 \alpha ^2 } \mathord{\left/ {\vphantom {{3a^2 \alpha ^2 } 2}} \right. \kern-\nulldelimiterspace} 2}} }}} \right. \kern-\nulldelimiterspace} {\sqrt {1 - {{3a^2 \alpha ^2 } \mathord{\left/ {\vphantom {{3a^2 \alpha ^2 } 2}} \right. \kern-\nulldelimiterspace} 2}} }}$, respectively. The horizons are given by the null-hyper surface condition $g^{\mu \nu } \left( {\partial _\mu  F} \right)\left( {\partial _\nu  F} \right) = 0$, where $F$ is the hyper-surface of black string. So, the black string has three horizons when $\Delta$ vanishes, the outer event horizon is located at $r_ +   = a^{ - 1} b^{{1 \mathord{\left/ {\vphantom {1 3}} \right. \kern-\nulldelimiterspace} 3}}$.
Now, applying the dragging coordinate transformation
\begin{eqnarray}
\label{eq4}
d\varphi  = d\phi  + \Omega dt= d\phi  + \frac{{r^2 \gamma \alpha ^4  - \Delta \gamma \delta \alpha ^2 }}{{r^2 \gamma ^2 \alpha ^4  - \Delta \delta ^2 }} dt,
\end{eqnarray}
one can rewrite the metric (\ref{eq2}) as
\begin{eqnarray}
\label{eq5}
ds^2  =  - A(r) dt^2  + \frac{1}{B(r)}dr^2  + C(r)d\varphi ^2  + D(r)dz^2 ,
\end{eqnarray}
where $A(r) = {{\Delta r^2 \left( {\alpha ^2 \gamma ^2  - \delta ^2 } \right)^2 } \mathord{\left/
 {\vphantom {{\Delta r^2 \left( {\alpha ^2 \gamma ^2  - \delta ^2 } \right)^2 } {\left( {\Delta \delta ^2  - \alpha ^4 r^2 \gamma ^2 } \right)}}} \right.
 \kern-\nulldelimiterspace} {\left( {\Delta \delta ^2  - \alpha ^4 r^2 \gamma ^2 } \right)}}$, $B (r) = \Delta ^{ - 1}$, $C(r) = r^2 \gamma ^2  - {{\Delta \delta ^4 } \mathord{\left/ {\vphantom {{\Delta \delta ^4 } {\alpha ^4 }}} \right. \kern-\nulldelimiterspace} {\alpha ^4 }}$, $D (r) = \alpha ^2 r^2$. On the event horizon, the $B(r)$ can be expended as $B(r)=B'(r_+)(r-r_+)+\mathcal{O}[(r-r_+)^2]$ \cite{ch17}. According to Ref.~\cite{ch14}, in order to study the tunneling behavior of massive bosons on the event horizon, one need to use the Proca equation
\begin{eqnarray}
\label{eq6}
D_\mu  \psi ^{ \nu \mu }  + \frac{{m^2 }}{{\hbar ^2 }}\psi ^\nu   = 0,
\end{eqnarray}
\begin{eqnarray}
\label{eq7}
\psi _{\nu \mu}  = D_\nu  \psi _\mu   - D_\mu  \psi _\nu   = \partial _\nu  \psi _\mu   - \partial _\mu  \psi _\nu  ,
\end{eqnarray}
where  $D_\mu$ are covariant derivatives, $\psi _\nu$ are depended on $\psi _t ,\psi _r ,\psi _\theta$ and  $\psi _\varphi$, and the mass of  bosons is $m$. The anti-symmetric tensor $\psi ^{\mu \nu }$ obeys the relation $\psi _{\mu \nu }  = \partial _\mu  \psi _\nu   - \partial _\nu  \psi _\mu$. Thus, with the help of equation $D_\mu  \psi _{\mu \nu }  = {{\partial _\mu  \left( {\sqrt { - g} \psi ^{\mu \nu } } \right)} \mathord{\left/ {\vphantom {{\partial _\mu  \left( {\sqrt { - g} \psi ^{\mu \nu } } \right)} {\sqrt { - g} }}} \right. \kern-\nulldelimiterspace} {\sqrt { - g} }}$, Eq.~(\ref{eq6}) becomes
\begin{align}
\label{eq8}
\frac{1}{{\sqrt { - g} }}\partial _\mu  \left( {\sqrt { - g} \psi ^{\nu \mu } } \right) + \frac{{m^2 }}{{\hbar ^2 }}\psi ^\nu   = 0.
\end{align}
where  $\psi ^\nu$  and $\psi ^{\mu \nu }$  are given as follows
\begin{align}
\label{eq9}
& \psi ^0  =  -{A ^{-1}}\psi _0 ,\psi ^1  = B\psi _1 ,\psi ^2  = {C ^{-1}}\psi _2 ,\psi ^3  = {D ^{-1}}\psi _3 ,\psi ^{01}  =  - {B}{A ^{-1}}\psi _{01},
 \nonumber \\
& \psi ^{02}  =  - {{(AC)} ^{-1}}\psi _{02} ,\psi ^{03}  =  - {{(AD)} ^{-1}}\psi _{03} ,\psi ^{12}  = {B}{C ^{-1}}\psi _{12} ,\psi ^{13}  = {B}{D ^{-1}}\psi _{13},
  \nonumber \\
& \psi ^{23}  = {{(CD)} ^{-1}}\psi _{23} .
\end{align}
Putting Eq.~(\ref{eq9}) into Eq.~(\ref{eq6}), and then, considering the relationship in Eq.~(\ref{eq7}), one obtains
\begin{align}
\label{eq10}
& \frac{1}{{\sqrt { - g} }}\left\{ {\partial _r \left[ {\sqrt { - g} \frac{B}{A}\left( {\partial _t \psi _1  - \partial _r \psi _{_0 } } \right)} \right] + \partial _\varphi  \left[ {\sqrt { - g} \frac{1}{{AC}}\left( {\partial _t \varphi _2  - \partial _\varphi   \psi _{_0 } } \right)} \right]} \right.
\nonumber \\
& \left. { + \partial _z  \left[ {\sqrt { - g} \frac{1}{{AD}}\left( {\partial _t \psi _3  - \partial _z  \psi _{_0 } } \right)} \right]} \right\} + \frac{{m^2 }}{{\hbar ^2 }}\frac{1}{A}\psi _0  = 0,
\end{align}
\begin{align}
\label{eq11}
& \frac{1}{{\sqrt { - g} }}\left\{ {\partial _t \left[ {\sqrt { - g} \frac{B}{A}\left( {\partial _t \psi _{_1 }  - \partial _r \psi _0 } \right)} \right] + \partial _\varphi  \left[ {\sqrt { - g} \frac{B}{C}\left( {\partial _r \psi _{_2 }  - \partial _\varphi  \psi _1 } \right)} \right]} \right.
 \nonumber \\
& \left. { + \partial _z \left[ {\sqrt { - g} \frac{B}{D}\left( {\partial _r \psi _{_3 }  - \partial _z \psi _1 } \right)} \right]} \right\} + \frac{{m^2 }}{{\hbar ^2 }}B\psi _1  = 0,
\end{align}
\begin{align}
\label{eq12}
& \frac{1}{{\sqrt { - g} }}\left\{ {\partial _t \left[ {\sqrt { - g} \frac{1}{{AC}}\left( {\partial _t \psi _{_2 }  - \partial _\varphi \psi _0 } \right)} \right] + \partial _r \left[ {\sqrt { - g} \frac{B}{C}\left( {\partial _\varphi  \psi _{_1 }  - \partial _r \psi _2 } \right)} \right]} \right.
  \nonumber \\
& \left. { + \partial _z \left[ {\sqrt { - g} \frac{1}{{CD}}\left( {\partial _\varphi  \psi _3  - \partial _z \psi _2 } \right)} \right]} \right\} + \frac{{m^2 }}{{\hbar ^2 }}\frac{1}{C}\psi _2  = 0,
\end{align}
\begin{align}
\label{eq13}
& \frac{1}{{\sqrt { - g} }}\left\{ {\partial _t \left[ {\sqrt { - g} \frac{1}{{AD}}\left( {\partial _t \psi _3  - \partial _z  \psi _0 } \right)} \right] + \partial _r \left[ {\sqrt { - g} \frac{B}{D}\left( {\partial _z  \psi _1  - \partial _r \psi _3 } \right)} \right]} \right.
  \nonumber \\
& \left. { + \partial _\varphi  \left[ {\sqrt { - g} \frac{1}{{CD}}\left( {\partial _z \psi _2  - \partial _\varphi  \psi _3 } \right)} \right]} \right\} + \frac{{m^2 }}{{\hbar ^2 }}\frac{1}{D}\psi _3  = 0.
\end{align}
In order to solve Eq.~(\ref{eq10})-Eq.~(\ref{eq13}), one can express the solutions to above equation set in the following form
\begin{eqnarray}
\label{eq14}
\psi _\nu   = \left( {c_0 ,c_1 ,c_2 ,c_3 } \right)\exp \left\{ {\frac{i}{\hbar }\left[ {S_0 \left( {t,r,\varphi ,z} \right) + \sum\limits_i {\hbar ^i } S_i \left( {t,r,\varphi ,z} \right)} \right]} \right\},
\end{eqnarray}
where $i=1,2,3,\cdots$. According to the WKB approximation and Eq.~(\ref{eq14}), Eq.~(\ref{eq10})-Eq.~(\ref{eq13}) can be reexpressed as
\begin{align}
\label{eq15}
& \left\{ {B\left[ { c_0 \left( {\partial _r S_0 } \right)^2 - c_1 \left( {\partial _t S_0 } \right)\left( {\partial _r S_0 } \right)  } \right] + C^{ - 1} \left[ c_0 \left( {\partial _\varphi  S_0 } \right)^2 - {c_2 \left( {\partial _t S_0 } \right)\left( {\partial _\varphi  S_0 } \right)  } \right]} \right.
  \nonumber \\
& \left. { + D^{ - 1} \left[  c_0 \left( {\partial _z S_0 } \right)^2  - {c_3 \left( {\partial _t S_0 } \right)\left( {\partial _z S_0 } \right) } \right]} \right\} + m^2 \psi _0  = 0,
\end{align}
\begin{align}
\label{eq16}
 &\left\{ {A^{ - 1} \left[ {c_0 \left( {\partial _t S_0 } \right)\left( {\partial _r S_0 } \right) - c_1 \left( {\partial _t S_0 } \right)^2 } \right] + C^{ - 1} \left[ {c_1 \left( {\partial _\varphi  S_0 } \right)^2  - c_2 \left( {\partial _t S_0 } \right)\left( {\partial _\varphi  S_0 } \right)} \right]} \right.
  \nonumber \\
 &\left. { + D^{ - 1} \left[ {c_1 \left( {\partial _z S_0 } \right)^2  - c_3 \left( {\partial _r S_0 } \right)\left( {\partial _z S_0 } \right)} \right]} \right\} + m^2 \psi _1  = 0,
\end{align}
\begin{align}
\label{eq17}
& \left\{ {A^{ - 1} \left[ {c_0 \left( {\partial _t S_0 } \right)\left( {\partial _\varphi  S_0 } \right) - c_2 \left( {\partial _t S_0 } \right)^2 } \right] + B\left[ {c_2 \left( {\partial _r S_0 } \right)^2  - c_1 \left( {\partial _r S_0 } \right)\left( {\partial _\varphi  S_0 } \right)} \right]} \right.
  \nonumber \\
& \left. { + D^{ - 1} \left[ {c_2 \left( {\partial _z S_0 } \right)^2  - c_3 \left( {\partial _z S_0 } \right)\left( {\partial _\varphi  S_0 } \right)} \right]} \right\} + m^2 \psi _2  = 0,
\end{align}
\begin{align}
\label{eq18}
& \left\{ {A^{ - 1} \left[ {c_0 \left( {\partial _t S_0 } \right)\left( {\partial _z S_0 } \right) - c_3 \left( {\partial _t S_0 } \right)^2 } \right] + B\left[ {c_3 \left( {\partial _r S_0 } \right)^2  - c_1 \left( {\partial _r S_0 } \right)\left( {\partial _z S_0 } \right)} \right]} \right.
  \nonumber \\
& \left. { + C^{ - 1} \left[ {c_3 \left( {\partial _\varphi  S_0 } \right)^2  - c_2 \left( {\partial _\varphi  S_0 } \right)\left( {\partial _z S_0 } \right)} \right]} \right\} + m^2 \psi _3  = 0.
\end{align}
By analysing the spacetime of the rotating black string, one can find 3 Killing vectors. Therefore, the action $S_0$ can be expressed as
\begin{eqnarray}
\label{eq19}
S_0  =  - \left( {\omega  - j\Omega } \right)t + W\left( r \right) + j\varphi  + \mathcal{Z} \left( z \right),
\end{eqnarray}
with the angular momentum $\omega$ and the energy of massive bosons $j$, respectively. Now, substituting  Eq.~(\ref{eq19}) into Eq.~(\ref{eq15})-Eq.~(\ref{eq18}), a $4\times4$ matrix is obtained as
\begin{eqnarray}
\label{eq20}
\Lambda \left( {c_0 ,c_1 ,c_2 ,c_3 } \right)^T  = 0.
\end{eqnarray}
where $\Lambda $ is a $4\times4$ matrix, which components are given as follows
\begin{eqnarray}
\label{eq21}
\begin{array}{l}
 \Lambda _{00}  =  - B\left( {W'} \right)^2  - {C^{-1}}\left( {\partial _z \mathcal{Z} } \right)^2  - {D^{-1}}j^2  - m^2 ,\Lambda _{01}  =  - BW'\left( {\omega  - j\Omega } \right), \\
 \\
 \Lambda _{02}  =  - {C^{-1}}\left( {\omega  - j\Omega } \right)\left( {\partial _z \mathcal{Z} } \right),\Lambda _{03}  =  -{D^{-1}}\left( {\omega  - j\Omega } \right)j, \\
 \\
 \Lambda _{10}  =  - {A^{-1}}\left( {\omega  - j\Omega } \right)W',\Lambda _{11}  = m^2  - {A^{-1}}\left( {\omega  - j\Omega } \right)^2  + {C^{-1}}\left( {\partial _z \mathcal{Z} } \right)^2  - {D^{-1}}j^2 , \\
 \\
 \Lambda _{12}  = {C^{-1}}\left( {\omega  - j\Omega } \right)\left( {\partial _z \mathcal{Z} } \right),\Lambda _{13}  = {D^{-1}}jW',\Lambda _{20}  =  - {A^{-1}}\left( {\omega  - j\Omega } \right)\left( {\partial _z \mathcal{Z}} \right), \\
 \\
 \Lambda _{21}  = BW'\left( {\partial _z \mathcal{Z} } \right),\Lambda _{22}  = {A^{-1}}\left( {\omega  - j\Omega } \right)^2  + B\left( {W'} \right)^2  + {D^{-1}}j^2  + m^2 , \\
 \\
 \Lambda _{23}  =  - {j}{D ^{-1}}\left( {\partial _z Z} \right),\Lambda _{30}  =  - {A^{-1}}\left( {\omega  - j\Omega } \right)j,\Lambda _{31}  =  - BW'j, \\
 \\
 \Lambda _{32}  =  - {C^{-1}}\left( {\partial _z \mathcal{Z} } \right)j,\Lambda _{33}  = m^2  -{A^{-1}}\left( {\omega  - j\Omega } \right)^2  + B\left( {W'} \right)^2  + {C^{-1}}\left( {\partial _z \mathcal{Z} } \right)^2 . \\
 \end{array}
\end{eqnarray}
In above equation, we denote that $W' = \partial _r W$. For obtaining the nontrivial solution of Eq.~(\ref{eq20}), it requires the  ${\rm{det}}\left( \Lambda  \right) = 0$. Thus, one has
\begin{eqnarray}
\label{eq22}
W_ \pm   =  \pm \int {\sqrt {\frac{{CD\left( {\omega  - j\Omega } \right)^2  - ACj^2  - ACDm^2  - AD\left( {\partial _z \mathcal{Z}} \right)^2 }}{{ABCD}}} dr},
\end{eqnarray}
and the above equation is computed as
\begin{eqnarray}
\label{eq23}
W_ \pm   =  \pm i\pi \left[ {\frac{{\left( {\omega  - j\Omega _{r_ +  } } \right)\gamma \alpha ^4 r_ + ^2 }}{{\left( {2\alpha ^4 r_ + ^3  + b} \right)\left( {\alpha ^2 \gamma ^2  - \delta ^2 } \right)}}} \right]+ \mathcal{O}(RealPart),
\end{eqnarray}
where $+/-$  represents the outgoing/incoming solutions on the outer event horizon. Here we only need the imaginary part since the real part of Eq.~(\ref{eq23}) does not contribute to the tunneling rate. The tunneling rate of massive bosons from 4-dimensional rotating black string is given by
\begin{eqnarray}
\label{eq24}
\Gamma  = \frac{{\Gamma _{\left( {emission} \right)} }}{{\Gamma _{\left( {absorpation} \right)} }} = \frac{{\exp \left( { - 2{\mathop{\rm Im}\nolimits} W_ +   - 2{\mathop{\rm Im}\nolimits} \Xi } \right)}}{{\exp \left( { - 2{\mathop{\rm Im}\nolimits} W_ -   - 2{\mathop{\rm Im}\nolimits} \Xi } \right)}} = \exp \left[ { - \frac{{4\pi \left( {\omega  - j\Omega _{r_ +  } } \right)\gamma \alpha ^4 r_ + ^2 }}{{\left( {2\alpha ^4 r_ + ^3  + b} \right)\left( {\alpha ^2 \gamma ^2  - \delta ^2 } \right)}}} \right].
\end{eqnarray}
Comparing Eq.~(\ref{eq24}) with the expression of Boltzman factor, that is $\Gamma  = \exp \left( {{E \mathord{\left/ {\vphantom {E T}} \right.
 \kern-\nulldelimiterspace} T}} \right)$ with the temperature $T$ and the energy of particles $\omega$ \cite{ch21}, the Hawking temperature of 4-dimensional rotating black string is
\begin{eqnarray}
\label{eq25}
T_H  = \frac{{\left( {2\alpha ^4 r_ + ^3  + b} \right)\left( {\alpha ^2 \gamma ^2  - \delta ^2 } \right)}}{{4\pi \gamma \alpha ^4 r_ + ^2 }}.
\end{eqnarray}
From Eq.~(\ref{eq24}) and Eq.~(\ref{eq25}), one can find that the tunneling rate and Hawking temperature of massive bosons from 4-dimensional rotating black string are related to the outer event horizon $r_+$, the mass $M$ and angular momentum  $J$ of the black string .  When comparing our results with those of scalars and fermions cases, it finds that they are the same \cite{ch20}.

\section{Massive bosons tunneling from the 5-dimensional black string}
\label{III}
In this section, the massive bosons tunneling behavior from  5-dimensional black string will be investigated by using the Proca equation. The metric of 5-dimensional black string is given by \cite{ch19}
\begin{eqnarray}
\label{eq26}
ds^2  =  - A\left( r \right)dt^2  + \frac{1}{{B\left( r \right)}}dr^2  + C\left( r \right)d\theta ^2  + D\left( r \right)d\varphi ^2  + E dz^2 .
\end{eqnarray}
where $F\left( r \right) = 1 - {{r_H } \mathord{\left/ {\vphantom {{r_H } r}} \right. \kern-\nulldelimiterspace} r}$, $G\left( r \right) =( 1 - {{r_H } \mathord{\left/ {\vphantom {{r_H } r}} \right. \kern-\nulldelimiterspace} r})^{ - 1}$, $C\left( r \right) = r^2$, $D\left( r \right) = r^2 \sin ^2 \theta$, $E=1$. $r_H$ is the event horizon of the 5-dimensional black string, we can rewrite the $B(r)$ as $B(r)=B'(r_+)(r-r_+)+\mathcal{O}[(r-r_+)^2]$. According to the line element of the 5-dimensional black string, one yields the following expression
\begin{align}
\label{eq27}
& \psi ^0  =  - {A ^{-1}}\psi _0 ,\psi ^1  = B\psi _1 ,\psi ^2  = {C ^{-1}}\psi _2 ,\psi ^3  = {D ^{-1}}\psi _3 ,\psi ^4  = {E ^{-1}}\psi _4 ,
\nonumber \\
& \psi ^{01}  =  - {B ^{-1}}{A}\psi _{01} ,\psi ^{02}  =  - {{(AC) ^{-1}}}\psi _{02} ,\psi ^{03}  =  - {{(AD)}^{-1}}\psi _{03} ,
\nonumber \\
&\psi ^{04}  =  - {{(AE)}^{-1}}\psi _{04}, \psi ^{12}  = {B^{-1}}{C}\psi _{12} ,\psi ^{13}  = {B ^{-1}}{D}\psi _{13} ,\psi ^{14}  = {B ^{-1}}{E}\psi _{14} ,
\nonumber \\
&\psi ^{23}  = {{(CD)}^{-1}}\psi _{23} , \psi ^{24}  = {{(CE)}^{-1}}\psi _{24} , \psi ^{34}  = {{(DE)}^{-1}}\psi _{34} .
\end{align}
Putting Eq. (\ref{eq27}) and Eq. (\ref{eq14}) into Eq. (\ref{eq6}), and keeping the first order term of $\hbar$, one has
\begin{align}
\label{eq28}
& B\left[ {c_1 \left( {\partial _t S_0} \right)\left( {\partial _r S_0} \right) - c_0 \left( {\partial _r S_0} \right)^2 } \right] + {C ^{-1}}\left[ {c_2 \left( {\partial _t S_0} \right)\left( {\partial _\theta  S} \right) - c_0 \left( {\partial _\theta  S_0} \right)^2 } \right]
\nonumber \\
&  + {D ^{-1}}\left[ {c_3 \left( {\partial _t S_0} \right)\left( {\partial _\varphi  S_0} \right) - c_0 \left( {\partial _\varphi  S_0} \right)^2 } \right] + {E ^{-1}}\left[ {c_4 \left( {\partial _t S_0} \right)\left( {\partial _z S_0} \right) - c_0 \left( {\partial _z S_0} \right)^2 } \right]
\nonumber \\
&  - m^2 \psi _0  = 0,
\end{align}
\begin{align}
\label{eq29}
& {A ^{-1}}\left[ {c_0 \left( {\partial _r S_0} \right)\left( {\partial _t S_0} \right) - c_1 \left( {\partial _t S_0} \right)^2 } \right] + {C ^{-1}}\left[ {c_1 \left( {\partial _\theta  S_0} \right)^2  - c_2 \left( {\partial _r S_0} \right)\left( {\partial _\theta  S_0} \right)} \right]
\nonumber \\
&  + {D ^{-1}}\left[ {c_1 \left( {\partial _\varphi  S_0} \right)^2  - c_3 \left( {\partial _r S_0} \right)\left( {\partial _\varphi  S_0} \right)} \right] + {E ^{-1}}\left[ {c_1 \left( {\partial _z S_0} \right)^2  - c_4 \left( {\partial _r S_0} \right)\left( {\partial _z S_0} \right)} \right]
\nonumber \\
&  + m^2 \psi _1  = 0,
\end{align}
\begin{align}
\label{eq30}
& {A ^{-1}}\left[ {c_0 \left( {\partial _\theta  S_0} \right)\left( {\partial _t S_0} \right) - c_2 \left( {\partial _t S_0} \right)^2 } \right] + B\left[ {c_2 \left( {\partial _r S_0} \right)^2  - c_1 \left( {\partial _\theta  S_0} \right)\left( {\partial _r S_0} \right)} \right]
\nonumber \\
&  + {D ^{-1}}\left[ {c_2 \left( {\partial _\varphi  S_0} \right)^2  - c_3 \left( {\partial _\varphi  S_0} \right)\left( {\partial _\theta  S_0} \right)} \right] + {E ^{-1}}\left[ {c_2 \left( {\partial _z S_0} \right)^2  - c_4 \left( {\partial _\theta  S_0} \right)\left( {\partial _z S_0} \right)} \right]
\nonumber \\
&  + m^2 \psi _2  = 0,
\end{align}
\begin{align}
\label{eq31}
& {A ^{-1}}\left[ {c_0 \left( {\partial _\varphi  S_0} \right)\left( {\partial _t S_0} \right) - c_3 \left( {\partial _t S_0} \right)^2 } \right] + B\left[ {c_3 \left( {\partial _r S_0} \right)^2  - c_1 \left( {\partial _r S_0} \right)\left( {\partial _\varphi  S_0} \right)} \right]
\nonumber \\
&  + {C ^{-1}}\left[ {c_3 \left( {\partial _\theta  S_0} \right)^2  - c_2 \left( {\partial _\theta  S_0} \right)\left( {\partial _\varphi  S_0} \right)} \right] +{E ^{-1}}\left[ {c_3 \left( {\partial _z S_0} \right)^2  - c_4 \left( {\partial _z S} \right)\left( {\partial _\varphi  S} \right)} \right]
\nonumber \\
&  + m^2 \psi _3  = 0,
\end{align}
\begin{align}
\label{eq32}
& {A ^{-1}}\left[ {c_0 \left( {\partial _z S_0} \right)\left( {\partial _t S_0} \right) - c_4 \left( {\partial _t S_0} \right)^2 } \right] + B\left[ {c_4 \left( {\partial _r S_0} \right)^2  - c_1 \left( {\partial _r S_0} \right)\left( {\partial _z S_0} \right)} \right]
\nonumber \\
&  + {C ^{-1}}\left[ {c_4 \left( {\partial _\theta  S_0} \right)^2  - c_2 \left( {\partial _\theta  S_0} \right)\left( {\partial _z S_0} \right)} \right] + {D ^{-1}}\left[ {c_4 \left( {\partial _\varphi  S_0} \right)^2  - c_3 \left( {\partial _\varphi  S_0} \right)\left( {\partial _z S_0} \right)} \right]
\nonumber \\
&  + m^2 \psi _4  = 0.
 \end{align}
Considering properties of spacetime, the action for 5-dimensional black string can be expressed as
\begin{eqnarray}
\label{eq33+}
S_0 =  - \omega t + W\left( r \right) + \Theta \left( \theta  \right) + j\varphi  + \mathcal{Z} \left( z \right),
\end{eqnarray}
where $\omega$  and  $j$ are the energy and angular momentum of the massive bosons. Inserting Eq.~(\ref{eq33+}) into Eq.~(\ref{eq28})-Eq.~(\ref{eq32}), the $5\times5$ matrix equation can be expressed as
\begin{align}
\label{eq34+}
\Lambda \left( {c_0 ,c_1 ,c_2 ,c_3 ,c_4 } \right)^T  = 0,
\end{align}
where d the elements of $\Lambda$ are
\begin{eqnarray}
\label{eq33}
\begin{array}{l}
 \Lambda _{00}  =  - B\left( {W'} \right)^2  - {C ^{-1}}\left( {\partial _\theta  \Theta } \right)^2  -{D ^{-1}}j^2  - m^2  + {E ^{-1}}\left( {\partial _z \mathcal{Z} } \right)^2 ,\Lambda _{01}  =  - B\left( {W'} \right)\omega , \\
 \\
 \Lambda _{02}  =  - {C ^{-1}}\omega \left( {\partial _\theta  \Theta } \right),\Lambda _{03}  =  - {D ^{-1}}\omega j,\Lambda _{04}  =  - {D ^{-1}}\omega \left( {\partial _z \mathcal{Z} } \right),\Lambda _{10}  =  - {A ^{-1}}\omega W', \\
 \\
 \Lambda _{11}  = m^2  - {A ^{-1}}\omega ^2  + {C ^{-1}}\left( {\partial _\theta  \Theta } \right)^2  - {D ^{-1}}j^2  + {K ^{-1}}\left( {\partial _z \mathcal{Z} } \right)^2 ,\Lambda _{12}  = {C ^{-1}}\omega \left( {\partial _\theta  \Theta } \right), \\
 \\
 \Lambda _{13}  = {D ^{-1}}jW', \Lambda _{14}  =  - {D ^{-1}}\left( {\partial _z \mathcal{Z} } \right)W',\Lambda _{20}  =  - {A ^{-1}}\omega \left( {\partial _\theta  \Theta } \right),\Lambda _{21}  = BW'\left( {\partial _\theta  \Theta } \right), \\
 \\
 \Lambda _{22}  =  - {A ^{-1}}\omega ^2  + B\left( {W'} \right)^2  + {D ^{-1}}j^2  + m^2  + {E ^{-1}}\left( {\partial _z \mathcal{Z} } \right)^2 ,\Lambda _{23}  =  - {D ^{-1}}j\left( {\partial _\theta  \Theta } \right), \\
 \\
 \Lambda _{24}  = {E ^{-1}}j \left( {\partial _\theta  \Theta } \right),\Lambda _{30}  =  - {A ^{-1}}\omega j,\Lambda _{31}  =  - B W'j,\Lambda _{32}  =  - {C ^{-1}}\left( {\partial _\theta  \Theta } \right)j, \\
 \\
 \Lambda _{33}  = m^2  - {A ^{-1}}\omega ^2  + B\left( {W'} \right)^2  + {C ^{-1}}\left( {\partial _\theta  \Theta } \right)^2  + {E ^{-1}}\left( {\partial _z \mathcal{Z} } \right)^2 ,\Lambda _{34}  =  - {D ^{-1}}\left( {\partial _z \mathcal{Z} } \right)j, \\
 \\
 \Lambda _{40}  =  - {A ^{-1}}\omega \left( {\partial _z \mathcal{Z} } \right),\Lambda _{41}  =  - BW'\left( {\partial _z \mathcal{Z} } \right),\Lambda _{42}  =  - {C ^{-1}}\left( {\partial _\theta  \Theta } \right)\left( {\partial _z \mathcal{Z} } \right), \\
 \\
 \Lambda _{43}  =  - {D ^{-1}}j\left( {\partial _z \mathcal{Z} } \right),\Lambda _{44}  = m^2  - {A ^{-1}}\omega ^2  + B\left( {W'} \right)^2  + {C ^{-1}}\left( {\partial _\theta  \Theta } \right)^2  + {D ^{-1}}j^2 , \\
 \end{array}
\end{eqnarray}
For obtaining a nontrivial solution, the determinant of the matrix $\Lambda$ must equals to zero. Therefore, $\det \left( \Lambda  \right) = 0$ leads to the following equation
\begin{align}
\label{eq34}
{\mathop{\rm Im}\nolimits} W_ \pm' &  =  \pm \rm{Im} \int {\sqrt {\frac{{C\left[ {DE\omega ^2  - AEj^2  - ADEm^2  - AD\left( {\partial _z \mathcal{Z} } \right)^2 } \right] - ADE\left( {\partial _\theta  \Theta } \right)^2 }}{{ABCDE}}}  dr}
\nonumber \\
& =  \pm i\pi \omega r_H,
\end{align}
where the plus/minus denote the outgoing/incoming solutions of massive bosons. Here, we need to neglect the real part of above equation. Finally, the tunneling rate of the massive bosons is
\begin{eqnarray}
\label{eq35}
\Gamma  = \frac{{\Gamma _{\left( {emission} \right)} }}{{\Gamma _{\left( {absorpation} \right)} }} = \frac{{\exp \left( { - 2{\mathop{\rm Im}\nolimits} W_ +   - 2{\mathop{\rm Im}\nolimits} \Xi } \right)}}{{\exp \left( { - 2{\mathop{\rm Im}\nolimits} W_ -   - 2{\mathop{\rm Im}\nolimits} \Xi } \right)}} = \exp \left( { - {{4\pi }}\pi \omega r_H } \right).
\end{eqnarray}
According to the the Boltzman factor, the Hawking temperature of 5-dimensional black string becomes
\begin{eqnarray}
\label{eq36}
T_H ' = \frac{1}{{4\pi r_H }}.
\end{eqnarray}
Obviously, the Hawking temperature of 5-dimensional black string is only depended on the event horizon $r_H$. If people investigates the scalar particles and fermions tunneling from the 5-dimensional black string, they would obtain the same results.

\section{Discussion and conclusion}
\label{Dis}
In this paper, the tunneling behaviors of massive bosons from the event horizon of 4-dimensional rotating black string and 5-dimensional black string have been investigated via the Proca equation. By analyzing Eq.~(\ref{eq24}) and Eq.~(\ref{eq25}), one can see that the tunneling rate and Hawking temperature of 4-dimensional rotating black string are related to the the event horizon, mass and angular momentum line densities of the rotating black string, and the negative cosmological constant. For the 5-dimensional black string case, it finds that $\Gamma'$ and $T_H'$ are only dependent on the the event horizon of the black string. Moreover, when comparing our results with those of scalars and fermions tunneling cases, it finds that they are the same. Meanwhile, according to the WKB approximation, we derived the Hamilton-Jacobi ansatz from Proca equation, which indicates that the Hamilton-Jacobi equation is a fundamental equation in the semiclassical theory. Therefore, Hamilton-Jacobi ansatz  can help us to investigate the semiclassical dynamic behavior of emit particles on the black holes' event horizon.

\vspace*{3.0ex}
{\bf Acknowledgements}
\vspace*{1.0ex}
This work is supported by the Natural Science Foundation of China (Grant No. 11573022).

\end{document}